# SECURITY PROPERTIES IN AN OPEN PEER-TO-PEER NETWORK

Jean-François Lalande, David Rodriguez, Christian Toinard

Laboratoire d'Informatique Fondamentale d'Orléans

Université d'Orléans – Ensi de Bourges

88 Bd Lahitolle, 18000 Bourges, France

jean-francois.lalande@ensi-bourges.fr

david.rodriguez@ensi-bourges.fr

christian.toinard@ensi-bourges.fr

*Abstract*

*This paper proposes to address new requirements of confidentiality, integrity and availability properties fitting to peer-to-peer domains of resources. The enforcement of security properties in an open peer-to-peer network remains an open problem as the literature have mainly proposed contribution on availability of resources and anonymity of users. That paper proposes a novel architecture that eases the administration of a peer-to-peer network. It considers a network of safe peer-to-peer clients in the sense that it is a commune client software that is shared by all the participants to cope with the sharing of various resources associated with different security requirements. However, our proposal deals with possible malicious peers that attempt to compromise the requested security properties. Despite the safety of an open peer-to-peer network cannot be formally guaranteed, since a end user has privileges on the target host, our solution provides several advanced security enforcement. First, it enables to formally define the requested security properties of the various shared resources. Second, it evaluates the trust and the reputation of the requesting peer by sending challenges that test the fairness of its peer-to-peer security policy. Moreover, it proposes an advanced Mandatory Access Control that enforces the required peer-to-peer security properties through an automatic projection of the requested properties onto SELinux policies. Thus, the SELinux system of the requesting peer is automatically configured with respect to the required peer-to-peer security properties. That solution prevents from a malicious peer that could use ordinary applications such as a video reader to access confidential files such as a video requesting fee paying. Since the malicious peer could try to abuse the system, SELinux challenges and traces are also used to evaluate the fairness of the requester. That paper ends with different research perspectives such as a dedicated MAC system for the peer-to-peer client and honeypots for testing the security of the proposed peer-to-peer infrastructure.*

## 1. INTRODUCTION

Historically, peer to peer softwares have been used to exchange files in order to bypass the limitations of centralized solutions. New problems arise when using peer to peer clients. First, the usage of these software are commonly associated to the idea that these exchanges are illegals, regarding the copyright violation [16]. If a user publishes a copyrighted file, the peer-to-peer system cannot prevent the spread of this file. Second, the peers are supposed to respect security requirements in order to guarantee some fairness in the network. For example, a peer is not supposed to download files without sharing its own files.

Security properties, i.e. confidentiality, integrity, availability are partially addressed in the literature. The focus is limited to the availability and confidentiality properties. The existing peer-to-peer architectures try to hide the user's activity in order to obtain a good level of anonymity. Moreover, the resources must remain available in the peer-to-peer network even if peers are faulty or try to abuse the protocol. On the contrary, the integrity property is poorly addressed: most of the papers considers integrity as the ability to transfer a file and to guaranty that it is identical to the original resource. But in practice, the user needs to specify how they want to share the resources in a more flexible way. Moreover, different security policies are required for various domains of usage.

This paper proposes a way to specify and enforce a flexible policy in order to allow the users to manage the required security properties in a peer-to-peer network. It covers properties such as confidentiality (the resource of a community is shared and stays in this community), integrity (some resource that needs integrity must not be modified even if shared between users) and availability (the resources that must be available must be spread over the peers). A new language is proposed for formalizing the requested security properties by introducing the notion of domains. That language enables to define the security policies of the different peers that can formally express their goals about the various resources they have. When conflicts appear between peers about the policies they have, a decision is taken by the resource's holder to send or not the resource to the requester. This paper shows how the decision can be taken using the analysis of the peer's policies.

If we consider a peer-to-peer network where all the peers honor the protocol and respect the security policies, the properties attached to resources will be respected even if the resources are exchanged in the network. Of course, a peer-to-peer network could have malicious peers that can try to abuse the protocol, which is the main issue that is treated in the literature and described in section 2. As our proposal is based on the negotiation of policies, the malicious peers could claim a policy they do not respect or ensure. Solutions for this problem are presented. It uses a real operating system security enforcement and a distant evaluation of the conformity of the claimed policy using various challenges and logs for both the peer-to-peer client and the operating system.

A prototype of a peer-to-peer Java client have been implemented using the JXTA technology. It implements our negotiation protocol and the enforcement of the requested policy through an automatic configuration of the target SELinux system. Simulations of our negotiation protocol is presented and real use cases of our peer-to-peer client are given.

## 2. STATE OF THE ART

The ancestors of distributed peer-to-peer systems are solutions with a centralized index of published resources. A peer can request the server in order to find resources owned by a peer it does not know. Then, the shared resource is exchanged from the identified source to the requesting peer. Other solutions with clusters of servers increase the reliability of the solution as in Kazaa networks. The next generation of peer-to-peer systems are totally distributed [13]. The index of published resources is ensured by the peers themselves [3, 6]. This avoids the reliability problem of the older peer-to-peer systems but introduces a loss of control on the publication and the exchanges. Therefore, a strong effort has been done to hide activities of these peer-to-peer systems [1, 5, 6, 10], but it is a very limited notion of security. This section presents how the security is addressed in the literature of peer to peer systems.

### 2.1. Confidentiality and Anonymity

The confidentiality property have been interpreted in different ways in the literature. The first effort in the first peer-to-peer networks was focused on anonymity of users [6, 7, 12], mainly to escape censorship or to be protected when downloading copyrighted resources. This anonymity is enforced by the protocol that allows the content of the message request to be encrypted to intermediary nodes. Several versions of this protocol [10, 6] have been proposed and are derivations of Chaum's mix protocol [4]. Another benefit of the encryption of the data is the integrity of the messages that can be verified by the recipient peer.

The use of relays can help the intermediary nodes for their own anonymity. As they acts as relays, they can hide their own activities inside the relayed requests. It becomes impossible to know if a node is acting for himself or for another nodes. This is called hiding or anti-censorship system [5]. The classical countermeasure in case of a tentative of hiding is to analyze the traffic flow of a node. An Internet Service Provider could try to analyze the IP packets in order to build a profile of the exchanged data [8]. The analysis could use information such as used ports, TCP or UDP protocols, the amount of flow that is exchanged [18].

## 2.2. Availability

Another classical property that a peer-to-peer system tries to guarantee is the availability of the resources, mainly in case of fail stop fault. The technical solutions for this issue rely on data replication: for example in GNUNet [11], the addresses and resources are duplicated on the neighbors of the node responsible of the resources. The resource's ownership is based on the hashcode of the file: the peers are organized in a binary tree and the file will be stored in the leaf whose binary number is the closest of the hashcode. Then, the neighbourhood will be the closest nodes of this leaf in the tree [3]. When a request is addressed to the node that is known as the closest node of the resource, it will eventually forward the request to a node it knows closest. The requester obtains a pool of peers that are closed to the resource. This pool guarantees a good availability of the resource even in case of failure or disconnection.

Other peer to peer file systems use different kind of DHT (Distributed hash table) like Chord [9] that uses a ring distribution of peers or more sophisticated organizations like in Can [17] that uses a multidimensional space. These solutions ensure the same kind of availability. Performance storage or resource localization as well as management of files and their issues are outside the scope of this paper.

## 2.3. Integrity

Several definitions of the integrity property exists. In [15] a comparison is done on the different definitions usually used. In peer-to-peer systems the integrity property can be interpreted as two different requirements:

- the concept of data quality: the resources must meet or exceed the quality expected by the user.
- the prevention of unauthorized modification of data.

For the data quality, the first peer-to-peer systems used metadata such as test comments or tags reported by the users. These reports are difficult to evaluate and easily exploited by a malicious node, even if sophisticated correlation algorithms try to aggregate comments to identify the good resources and to exclude fakes.

For the integrity of the exchanged data, the protocols often combines anonymity and verification when using cryptographic keys to encrypt the requests and responses. One of the most known protocol that provides integrity of messages in ECRS [14] which is a variant of the CHK encoding scheme used in Freenet. The principle of ECRS is to provide a way to encrypt the response that will return from an unknown node to the initiator of the request. This way, the node that sent the request can check that the node that answered as really a resource that matches the request and are not a malicious node that have intercepted the request and altered the information.

Other peer-to-peer systems have been proposed to manage the modification of a resource (OceanStore, Ivy and Pastis). It can be seen as a sort of access control systems that allows the publisher of the resource to control who is authorized to update the resource. In Pastis, the owner can delegate the write permission to a group of peers. These proposals are the first steps of the integrity property. The frameworks are complex to deploy as they rely on cryptographic keys and signatures that have to be deployed after having authenticated the participating peer-to-peer nodes.

### 2.4. Discussion

Existing solutions suffer from several major limitations :

- opened peer-to-peer networks cannot be controlled. Indeed, each peer has local root privileges that authorize all the possible resource or protocol compromising. Thus, each peer can abuse the others about the enforcement of the security that it offers to the others.

- they do not address advanced security properties but only limited ones such as anonymity, confidentiality and integrity of messages. Integrity, confidentiality and availability of domains is missing. Moreover, relationships between the domains is not addressed.

- definition of security properties is not permitted. Thus, a peer cannot announce the proposed security policy.

- evaluation of the enforcement of the proposed security policy is not supported. So, attempts to abuse the system cannot be detected.

- conflicts between distant security policies must be resolved in order to authorize the resource exchange.

- finally, one cannot find any model of trust that evaluates the reputation of a peer according to the requested security properties such as integrity and confidentiality of domains.

The following sections address those different limitations in order to propose a new model to express and enforce a larger range of security properties in relation with various domains of usage. Moreover, evaluation strategies proposed to cope with conflicting but also malicious peers.

### 3. SECURITY LANGUAGE

Each peer can have a great number of resources. To simplify their management, they are grouped together into domains to which are applied a set of security properties. The security policy of a peer is composed of all its domains and security properties.

### 3.1. Domains

A domain d is defined as a unity of organization that groups multiple resources which have the same set of security properties. The concept of domain is quite opened because it depends of the community that will define the domains. Our basic example is to consider three domains : a free domain, a fee paying domain, and a private company domain.

The free domain defines a group of resources that can be exchanged without any restriction. The fee paying domain defines the resources that need the payment of fees in order to have access to these resources. The private domain defines private medias. When a resource is published on the peer-to-peer network (it is the first time the resource is added), the resource is necessarily associated to a domain, as described later in section 3.4.

### 3.2. Operation on domains

A peer creates a domain when requested by the user. A domain will contain downloaded resources and is empty at the beginning. The user will discover the existence of a domain, for example on the web, and will ask to the peer-to-peer client to create the domain. The deletion of a useless domain can be requested by the user. In this case, all the resources of the domain should be dropped.

### 3.3. Security properties

Integrity, confidentiality and availability are general properties which need to be adapted to peer to peer networks. Indeed, they are fully opened, decentralized and sometimes anonymous, so it is a challenge to define and to insure a specific security property for a given resource. Our architecture supports two kinds of properties : negative rules and positive rules which are prohibition properties and permission properties.

### 3.3.1. Prohibition properties

Intuitively, a prohibition property can be "the resource must not change of domain" and a permission property "the resource is encouraged to change of domain". Prohibition properties are basically the removal of rights in the peer-to-peer network. By default, a request have to be honoured because the goal of a peer-to-peer network is to exchange files between users. A policy that promotes prohibition properties tries to reduce the freedom of the users, in order to guarantee properties that are conflicting with the freedom property.

***Confidentiality*** The confidentiality property expresses that the resources of a domain must be maintained in this domain over the peer-to-peer network. This is not the classical definition of the confidentiality property as it does not answer the question "who can access the resource over the peer-to-peer network ?". That means that our goal is not to restrict the access of the resource to a set of users (the users are not known in advance in an open peer-to-peer network). Our goal is to guarantee that the resources stay associated to the considered domain.

*confidentiality(d1)*: no resources can exit domain d1.

*confidentiality(o1)*: resource o1 must not exit its current domain.

For example, the property is useful for the fee paying and a private company domain, that want that their files stay in their respective domains:

```
confidentiality(fee_paying)
confidentiality(private_company_A)
```

Moreover, the confidentiality property can be less restrictive if considering two domains that exclude each other: confidentiality(d1, d2) : a resource of domain d1 must not be able to be written in domain d2. For example, the property is useful to express that the fee paying domain must not share resources that will reach the free domain:

```
confidentiality(fee_paying, free)
confidentiality(private_company_A, {free, fee_paying})
```

The confidentiality property does not prohibit a resource to be shared as long as it stays in the same domain. Next, we introduce a property that prohibit the sharing.

***No share property*** The no share property expresses the need of prohibiting the share of the resource with another peer on the peer-to-peer network. The resource can eventually change of domain on the same peer, but the peer must try to guaranty that the resource will not be sent to another peer.

*noshare(d1)*: the resources in domain d1 must not be shared anymore.

*noshare(o1)*: the resource o1 must not be shared anymore.

This property can be used when a resource is sent by peer A to peer B in order to request that peer B does not share that resource. In this case, peer A allows peer B to get the resource but request the noshare property from peer B:

> noshare(movie_file)

***Integrity*** The integrity property aims to guaranty that the resources of a domain will not be modified by a peer.

*integrity(d1)*: resources in domain d1 must not be modified.

*integrity(o1)*: resources o1 must not be modified.

For example, a PDF file is broadcasted over the peer-to-peer network and the author requires that the file will not be modified by any peer that gets a copy of it:

> integrity(report.pdf)

***No publication*** The no publication property aims to guaranty that no resource will be published into a given domain.

*nopublication(d1)*: no resource in domain d1 must be published.

For example, a peer can decide that the domain fee paying will only be used to download files and that it has no reason to publish files into this domain:

> nopublication(fee_paying)

### 3.3.2. Permission properties

The presented prohibition properties restrict the freedom of the users: the global policies of the resources will become more and more strict. The worst case is the no share property that forbid definitively a peer to exchange a resource. To counterbalance these rules that can lead the system in a freezed state, we introduce permission properties that can be seen as positive properties that help the users to exchange files. This is a way to reenforce the freedom of users to share resources with its community.

***Cooperation*** Two companies could be interested in setting up a strong cooperation between them. Each company can create a private domain and will probably use the confidentiality property in order to protect their own data. In contrast, two domains can cooperate if the resources can easily be shared between them.

*cooperation(d1, d2)*: all resources of domain d1 must be available for sharing into domain d2.

*cooperation(o1, d2)*: resource o1 must be available for sharing into domain d2.

For two companies A and B using private company A and private company B as domains, the peers of these companies can setup these properties to authorize any sharing of private data:

> cooperation(private company A, private company B)
> cooperation(private company B, private company A)

***Spread*** The spread property is a generalization of the cooperation property. The idea of spreading a resource is that the owner of the resource wants the resource to be spread as much as possible on the peer-to-peer network, copied into the same or other domains.

*spread(d1)*: all resources of domain d1 must be available as much as possible and shared with other domains.

*spread(o1)*: resource o1 must be available as much as possible and shared with other domains. For example, security patches could be spread on the domain kernel patch for the linux kernel:

```
spread(kernel_patch)
```

### 3.3.3. Operation on properties

There are two ways of updating security properties. A user can decide to add or remove a security property for a domain or a file. The second case is when the peer receives a request from the owner of a resource to modify its policy. These two situations can generate conflicts of properties.

### 3.3.4. Conflicting properties

Security properties can be added to a resource or to a domain. Each update can bring up a situation of conflicting properties. The table 1 shows if rule r1 can conflict with rule r2. In case of conflict, a decision must be taken to select one of the two rules or to perform more general modifications of the policy.

Table 1. Conflicting properties.

| r1 / r2 | Conf. | Integ. | Spread | No pub. | No share. | Coop. |
|---|---|---|---|---|---|---|
| Confidentiality | | | x | | | x |
| Integrity | | | | | | |
| Spread | x | | | | x | |
| No publication | | | | | | |
| No share | | | x | | | x |
| Cooperation | x | | | | x | |

## 3.4. Publication

The publication of a resource can be done by any peer in the system. When publishing a resource, the user chooses a domain where to drop it. The peer-to-peer client will check the policy in order to detect a possible violation, for example a nopublication rule concerning the targeted domain. When a user performs a publication, the domain's file is not published across the network but only the name of the file and its location are sent. The mechanism of storing and retrieving the resource are managed as in classical peer-to-peer networks and are not addressed in this paper.

Moreover, the peer will eventually add new security properties for the considered resource to the local policy. The publication will have this form:

```
publish(security properties, resource, targeted domain)
```

For example, when publishing a report, a user of company A will ask to the peer-to-peer client:

```
publish({confidentiality, integrity}, reportA.pdf, private_company_A)
```

The confidentiality property is already ensured by the properties of domain private company A in listing 1: there is no need to add another confidentiality property. For the integrity property, a new rule integrity(reportA.pdf) will be added on peer A to get a satisfying security policy.

```
Domains: free, fee paying, private_company_A
```

```
confidentiality(fee paying)
confidentiality(private company A)
integrity(reportA.pdf)
cooperation(private company A, private company B)
```

Listing 1. Domains and security properties of peer A.

## 4. PEER-TO-PEER NEGOTIATIONS

The described security language allows a peer-to-peer client to express the properties that he wants to apply on its resources. This is crucial for the resources that the peer wants to share on the network. But this language is also used when two peers are negotiating for an exchange. The principles of a negotiation between a peer A (pA), owner of a resource, and an asking peer B (pB) is described in this section.

### 4.1. Negotiation basics

First the peer pB sends a request to pA for a file on a target domain called dB. The asked resource is situated in a domain dA that could be different of dB. In fact, two situations are possible:

- The two peers know each other or have previously exchanged information about the domain they will use to share data. It means that both peers knows the string that will be used as domain's name, i.e. dA = dB.

- The two peers have never met before. It means that the peer pB does not know the name of the domain dA where is situated the resource r. As a consequence, dA ≠ dB.

If dA = dB, the peer B have probably applied the same policy to the domain dB, to be consistent with peer A. This way, the negotiation will have the best chance to succeed, as the peer A will have the guaranty that the policy applied to its resource will stay guaranteed when the resource will be exchanged with B. Nevertheless, even if the domain's names are the same, the policies applied to both domains could be different.

Thus, pA will ask pB to send the part of its policy that is currently applied to the domain dB. Indeed, the peer B as no reason to send its whole policy to peer A. The peer B does not want to reveal the rules that are applied to other domains he created. Only the rules applied to dB are needed in the negotiation because they will be associated to the resource after the exchange. The peer pA will be able to check if the received policy hurts the policy attached to the domain dA. For this purpose, the Table 1 is used to determine if one of the proposed property of domain db enters in conflicting with one of those of the domain dB.

A basic decision is to decline the request of exchange if any property of dB hurts a property of dA. This way, the peer pA is sure to preserve the properties on the resources of dA. If the domain dA as no property, the peer cannot hurts the policy of A and the peer pB is free to propose any new property on the domain pB. Thus, a peer has the possibility to make evolve its policy for the considered resource: each pear, that requests a resource, can add a new property, if this property does not enter in conflict with the previous associated properties. For example, if a property spread(d) is attached to a resource, a peer will not be able to add the property confidentiality(d) because the peer that owns the resource will refuse the transfer.

### 4.2. Negotiation with malicious peers

For properties that deal with security, we should evaluate if the proposal resists to attacks that try to defeat the security mechanisms. To defeat the mechanism, a malicious peer can announce any security properties to the other peers. Nevertheless, two cases must be distinguished:

- if the malicious peer have no information about the properties associated with the requested resource, it will probably fail negotiating with the owner because the proposal will hurt the owner policy.
- if the malicious peer knows precisely what policy is required for the requested resource, it can claim to ensure the same properties on the target domain, even if it will not honor those properties later on.

Obviously, in a general manner, one can never prevent a malicious peer to simulate a satisfying behaviour. However, simulating a correct behaviour becomes difficult when the malicious peer:

1. must guess the requested policy,
2. is submitted to challenges
3. is automatically configured to enforce the requested policy using a MAC system such as SELinux that protects that peer at the Operating System scale.

The next section describes how to compute challenges and evaluate the accuracy of the response.

### 4.3. Evaluating the fairness of a peer

#### 4.3.1. Evaluation of the proposed policy

The proposed policy could be evaluated regarding the possible conflicts with the requested policy. When the peer receives the proposed policy from the requester, it will check if the proposed policy is conflicting with each property of the proposed policy:

- -1 : the property is clearly violated by the policy;
- 0 : the property is not provided by the policy;
- 1 : the property is respected.

A basic scenario is to sum the evaluations to obtain a global score. If positive, the global evaluation is positive. It means that most of the properties are respected by the requester of the resource and that some of them may be violated.

#### 4.3.2. History

The peer that owns the requested resource might ask a part of the history file of the operations that the requester did. This is a way to check if the requester is respecting the policy it proclaims. This log file could also be a fake log file, but it is a difficult task to provide a fake log file that is coherent with all other parties in a distributed systems. For example, a malicious peer can simulate the existence and transactions with a fake peer but the peer that evaluates the history can check if the fake peer exists or has been viewed in the past. A proposal of evaluation of the history during a given time t is given below:

- -1 : at least one log record indicates that the policy has been violated during t;
- -2 : the property has been violated before t;
- 1 : no infraction has been viewed.

Again, these figures can be combined to get a global score about the evaluation of the history file of the evaluated peer. Evaluation of an history is reduced to the past. In order to have a real time view of the reliability of requester, challenges are described in the following section. As detailed in the sequel, our architecture enforces the requested peer-to-peer security properties using SELinux policies. So, the history also includes some SELinux traces such as the SELinux trace corresponding to the attempt to open a protected resource using a malicious application e.g. vim for opening a confidential file. Further details will be given in the experimentation section.

### 4.3.3. Challenges

These challenges are delegated to the peers that A trusts. Thus, it becomes more difficult for a malicious peer to answer accordingly to the various peers because it will try to be consistent with conflicting challenges. In order to improve the evaluation, a trusted peer C will first send conflicting requests to peer B in order to make it change its policy accordingly. For example, C will request the spread property that is conflicting with the confidentiality property requested by A for the same domain. Afterwards, C will send challenges to A for evaluating the response to the transfer challenges. The purpose of that paper is not to choose between the best way to evaluate the response to the challenge, since various trust measurements exist in the literature, but to propose an overall architecture that aims at reusing existing trust formulas to evaluate the response to challenges. In the sequel, simulation runs present the evaluation of the challenges carried out by the trusted distant peers.

Since our architecture enforces the requested peer-to-peer security properties using SELinux policies, the proposed challenges include some SELinux challenges such as the request for the distant peer to try to open a protected resource using a malicious application e.g. vim for opening a confidential file. Further explanation will be given in the sequel about those mechanisms.

### 4.3.4. Reputation

The peer A computes the evaluations, carried out during the challenge phase, in order to define a challenge for the requesting peer B. Thus, the various distributed challenges enable to compute an overall reputation for the requesting peer B. Various mathematical measurements from the literature can be reused to evaluate the reputation of a peer B. The purpose of that paper is not to choose between the best mathematical formulas to evaluate the reputation, but to give a global architecture that can reuse efficiently the existing reputation formulas. In the sequel, simulation of reputation measurements are described.

## 5. ENFORCING THE SECURITY PROPERTIES AT THE OS LEVEL

After the evaluation of the security policy presented in the previous section, the remaining security pitfalls deal with the local attacks carried out both at peer A and peer B sides. Let us assume a corrupted web navigator that could try to read the protected resources from the resources of a peer-to-peer client. In order to prevent from such attacks or security violations, the requested peer-to-peer security policy is projected onto a SELinux protection policy. Thus, the SELinux mandatory access control mechanism helps to enforce the security policy and prevents the web navigator to access to the protected resources but allows the peer-to-peer client to access it.

So, the following enforcement is provided:

- peer-to-peer domains are protected against malicious applications
- external applications, such as a video reader, are prevented from violating the requested peer-to-peer security policies, such as the confidentiality of video files.

### 5.1. Projection of the peer-to-peer security properties onto SELinux protection policies

A dedicated tool P2PtoSELinux has been developed to convert the peer-to-peer security properties into consistent SELinux policies. That tool takes a peer-to-peer security policy p2p_properties.xml as input and produces a SELinux policy pol_selinux as output.

The listing 2 shows an extract of a peer-to-peer policy file, defining a domain name called domainA associated to different security properties and a resource corresponding to the file /root/secret.txt that belongs to domain A. That resource has a dedicated property corresponding to the cooperation with another domain. This policy is represented on figure 1. The security properties of domain A are listed and the dedicated property is written inside the file. Each security property is linked to the domain that is concerned by the property.

```
<policy>
<domain id="1" name="domainA">
 <property type="confidentiality">
  <target domainid="3"/>
 </property>
 <property type="integrity"/>
 <property type="cooperation">
  <target domainid="4"/>
 </property>
</domain>
<file id="2" path="/root/secret.txt" domainid="1">
 <property type="cooperation">
  <target domainid="2"/>
 </property>
</file>
</policy>
```

Listing 2. P2P XML Policy.

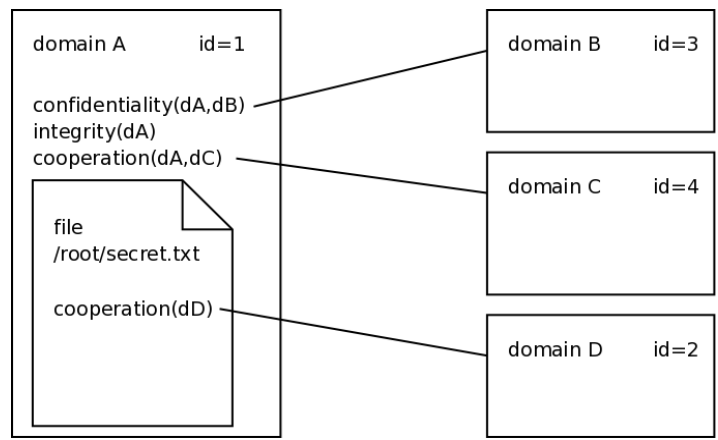

Figure 1: Policy representation for the XML policy of listing 2

The listing 3 shows an extract of the resulting SELinux protection policy produced by the P2PtoSELinux tool. As one can see, a default policy is computed but also dedicated policies for the various security properties i.e. confidentiality, integrity, no publication, no share, cooperation and spread. Those protection policies limit the permission for the corresponding resources. A peer-to-peer client will use the various corresponding security context according to the required domain. Thus, a confidential resource will be labelled with the Confidentiality SELinux context. Usually, the Confidentiality context cannot be read by any external application such as a video reader since it does not have the required privileges. However, a video reader can play a resource protected by the Confidentiality context if it has a satisfying SELinux subject context. In order for the end users to read confidential information, they must have the corresponding privileges and the application must transmit to a satisfying SELinux subject in order to be able to read that information.

```
Default:
allow file {read write unlink create append mounton rename lock execute geattr setattr}
allow dir {read write unlink search create mounton getattr setattr rename add_name remove_name reparent rmdir}
Confidentiality: Default +
neverallow file {read append setattr}
neverallow dir {read search setattr}
```

```
Integrity: Default +
neverallow file {write unlink append rename setattr}
neverallow dir {write unlink setattr rename remove_name rmdir}
No publication, NoShare: Default +
neverallow file {create setattr mounton}
neverallow dir {create setattr add_name remove_name rmdir mounton}
Cooperatoin, Spread: Default
```

Listing 3: SELinux rules

### 5.2. Obtained SELinux protection

The listing 4 shows the trace of a SELinux control associated with the editor vim attempting to read the file secret.txt that is protected by the peer-to-peer Confidentiality property. The SELinux context system_u:object_r:domainA_t associated to the secret.txt file prevents the vim application from the reading access.

So, that example shows that an external application such as vim cannot violate the corresponding peer-to-peer Confidentiality property. Thus, the proposed solution enforces the security of the requested properties within the local machine, since a malicious external application fails to compromise the peer-to-peer security properties for the local peer-to-peer resources.

```
audit(1229395253.757:369): avc:  denied  { read } for pid=4241 comm="vim"
name="secret.txt" dev=sda3 ino=179226 scontext=user_u:user_r:user_t
tcontext=system_u:object_r:domainA_t tclass=file
```

Listing 4: SELinux control

## 6. IMPLEMENTATION AND EXPERIMENTATION

We are currently working on the implementation of a peer-to-peer network which can answer the requirements of the model of section 3. A first Java simulation of the protocol and trust computation has been released. Moreover, a first implementation of a peer-to-peer client is currently available. The user interface will be presented in the second subsection in order to show the usage of the security properties. That interface is a front end for the P2PtoSElinux tool. Thus, the peer-to-peer security properties requested by the end user are automatically projected onto SELinux, protecting the target system from the corruption of the other applications.

### 6.1. Trust simulation

Listing 5 shows the simulation code that setup two peers JFL and David. David asks the file "contract" to peer JFL that have the confidentiality property on the "ensib" domain. David claims to put the file in the free domain which ensures the spread property.

Listing 6 shows the resulting negotiation between peer JFL and David. It shows the details of the computation of the Trust value (Tv) for the two required properties for the domain "ensib" against the David policy using its history and challenges: Tv(integrity,David) equals 0.38 and Tv(confidentiality,David) equals 0 for the domain "ensib". This last value forces JFL to refuse the resource exchange, because of the value of a fixed threshold (0.2).

```
// Creating peers, domains, files on JFL side
Domain ensib = new Domain("ensib");
Domain free = new Domain("free");
Resource firefox = new Resource("firefox");
Resource contract = new Resource("contract");
Peer jf = new Peer("JFL");
Peer pc1 = new Peer("C");
Peer pc2 = new Peer("C");
// Initiating the reputation of peer JF
jf.knows(pc1,0.8);
```

```
jf.knows(pc2,0.9);

// Adding resource in domains
jf.add(firefox, free);
jf.add(contract, ensib);
// Applying security properties on domains
Property confid = new Property("confidentiality");
Property integ = new Property("integrity");
Property coop = new Property("cooperation");
jf.add(confid, ensib);
jf.add(integ, ensib);
jf.add(coop, free);

// Creating peers, domains, files on David side
Domain free2 = new Domain("free");
Property spread = new Property("spread");
Peer david = new Peer("David");
david.add(spread,free2);

// Simulation
jf.affiche();
david.affiche();
david.ask(jf, "contract", "free");
```
Listing 5: Example of simulation code

```
 [Display JFL] <domain> ensib secured by [confidentiality, integrity]
 [Display JFL] <file> contract in ensib under [confidentiality, integrity]
 [Display JFL] <file> firefox in free under null
 [Display David] <domain> free secured by [spread]
David: I asks to peer JFL the file contract to be put in free
JFL: Peer David asking file contract
JFL: Peer David will put the file in domain free
JFL: File contract found.
JFL: File is in domain ensib
JFL: Security properties [confidentiality, integrity]
David: someone asking policy for domain free
David: returning policy [spread]
  (Eval) JFL Computation of Eval(David,confidentiality)
  (Eval) JFL Target domain has property spread
  (Eval) JFL the properties of David's free domain hurts the required property confidentiality
  (Eval) Eval(David,confidentiality)=-1
  (Eval) Hist(David,confidentiality)=2
  (Eval) Chal(confidentiality,David)=0.4
  (Eval) EvalHist(confidentiality,David)=0.0
  (Eval) Tv(confidentiality,David)=0.0
  (Eval) Peer refused (0.0<0.2) for confidentiality trust decreased to 0.48
  (Eval) JFL Computation of Eval(David,integrity)
  (Eval) JFL Target domain has property spread
  (Eval) Eval(David,integrity)=0
  (Eval) Hist(David,integrity)=2
  (Eval) Chal(integrity,David)=0.7266666666666667
  (Eval) EvalHist(integrity,David)=0.5
  (Eval) Tv(integrity,David)=0.38524635476491
  (Eval) Peer not fully trusted (0.2<0.38524635476491<0.5) for integrity trust decreased to 0.47
JFL: one of the property is refused: refusing request.
David: peer JFL REFUSED to send the file.
```
Listing 6: Example of negotiation

A second example is given in listings 7 and 8. This time, the resource firefox is asked by David that declares to put it in domain "free". As the cooperation property does not hurt the spread property, the computation of Tv(cooperation,David) gives 0.44 that is sufficient to allow the resource exchange (greater than the fixed threshold 0.2).

```
//david.ask(jf, "contract", "free");
david.ask(jf, "firefox", "free");
```
Listing 7: Firefox asked by David

```
[Display JFL] <domain> ensib secured by [confidentiality, integrity]
```

```
[Display JFL] <domain> free secured by [cooperation]
[Display JFL] <file> contract in ensib under [confidentiality, integrity]
[Display JFL] <file> firefox in free under [cooperation]
[Display David] <domain> free secured by [spread]
David: I asks to peer JFL the file firefox to be put in free
JFL: Peer David asking file firefox
JFL: Peer David will put the file in domain free
JFL: File firefox found.
JFL: File is in domain free
JFL: Security properties [cooperation]
David: someone asking policy for domain free
David: returning policy [spread]
  (Eval) JFL Computation of Eval(David,cooperation)
  (Eval) JFL Target domain has property spread
  (Eval) Eval(David,cooperation)=0
  (Eval) Hist(David,cooperation)=2
  (Eval) Chal(cooperation,David)=0.8500000000000001
  (Eval) EvalHist(cooperation,David)=0.5
  (Eval) Tv(cooperation,David)=0.44243254304683094
  (Eval) Peer not fully trusted (0.2<0.44243254304683094<0.5) for cooperation trust decreased to 0.49
JFL: request accepted.
David: peer JFL accepted to send the file.
```

Listing 8: Negociation for Firefox

### 6.2. Peer-to-peer client managing security properties

A peer-to-peer client has been developed using a JXTA module that provides peer-to-peer primitives (file transfer, connection of new clients). The figure 2 shows the graphical interface of the client. It allows to edit the security policy of the local domains and files. The lower left window shows the shared files and the domains associated to those files. The right side enables to change the security property associated to the file or to the domain. The upper window shows the distant shared files or the results of a specific string search.

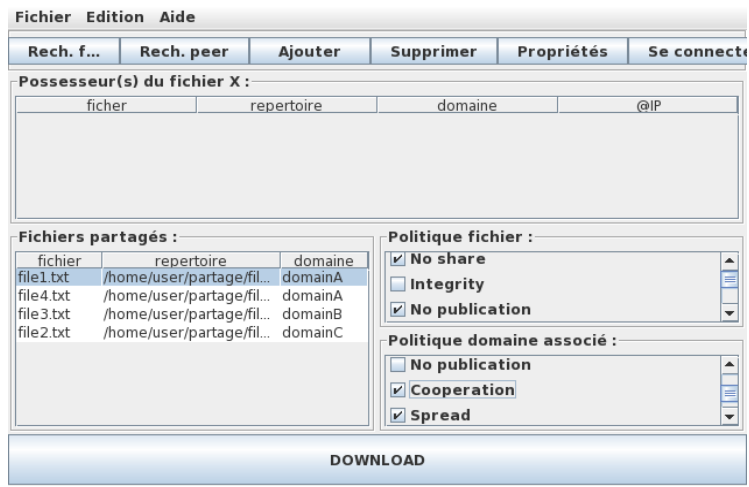

Figure 2. Graphical interface of the peer-to-peer client

Configuration of the requested security properties is easy through that user interface. Moreover, the trust algorithm is completely transparent to the end user.

### 6.3. P2PtoSELinux performances

The figure 3 shows how the execution time evolves according to the number of considered domains. For 10 domains, P2PtoSElinux takes 1.67 seconds to compute the corresponding SELinux policy. For 640 domains the execution time takes 2.736 seconds. So, the computation time is linear with the number of considered domains when considering more than 100 domains.

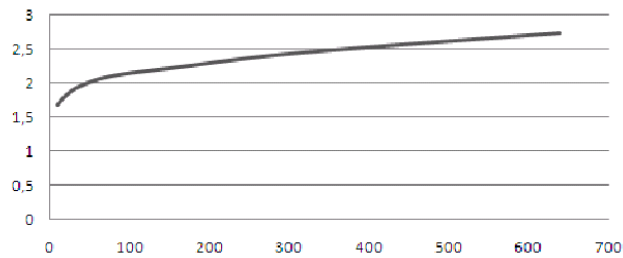

Figure 3: Execution time according to number of domains

### 6.4. SELinux challenges and traces

As presented previously our architecture uses SELinux traces and challenges to evaluate the trust and reputation of the requester. Let us give some examples of the produced challenges and traces associated with the target resource secret.txt. When a peer B requests a resource associated with domainA, peer B sends its peer-to-peer security policy corresponding to its target domain such as presented in listing 2. Thus, a trusted peer C sends to B a challenge such as defined in listing 9 in order to request the peer-to-peer client to execute the vim /root/secret.txt command using the subject SELinux context user_u:user_r:user_t. If the peer-to-peer client that is executed onto B is safe, it should answer with a failed SELinux trace such as presented in listing 4.

```
scontext=user_u:user_r:user_t vim /root/secret.txt
```

Listing 9. SELinux control

## 7. PERSPECTIVES

The enforcement of the proposed security properties brings several new perspectives.

First, conflicting security properties will bring conflicting SELinux policies, which is not acceptable for SELinux. A decision engine have to decide how to resolve those conflicting policies.

Second, a MAC mechanism could be integrated within the peer-to-peer client itself. That dedicated MAC mechanism would prevent from illegal flows inside a peer-to-peer client in order to deal with a corrupted peer-to-peer client. This MAC mechanism could be based on the JAAS module provided by the Java virtual machine that allows to control the access to the resources of the system such as the network or the files. Thus, two levels of access control could be setup: one fine grained access control could be applied at the software level, and one strongest access control at the OS level.

Finally, large scale experimental results are planned in order to evaluate the protocol's robustness. Different experiments will bet setup. First, a simulation of a large network of peers will permit to introduce a malicious peer that will violate the policy it claims. This will allow to analyze if the system behave well in a large system with thousands of peers. Second, experiments are planned based on our experience in high interaction honeypots [2] where compromised hosts are observed in order to analyze attackers behaviours. A peer-to-peer honeypot will be setup to invite attackers to try to get resources of the participants, protected by security properties. This experiment could help to improve the security of the proposed solution analysing the attacks in a real peer-to-peer network.

## ACKNOWLEDGEMENTS

This paper was supported by the work of several engineering students in computer sciences for their master degree from Ensi of Bourges, France: Virginie Klein, Fabien Le Solliec, David Rosa, Michael Souchet.

## REFERENCES

[1] Androutsellis-Theotokis, S. & Spinellis, D. (2004) "A survey of peer-to-peer content distribution technologies", ACM *Computing Surveys (CSUR)*, Vol. 36, No. 4, pp. 335-371.

[2] Briffaut, J., Lalande, J.-F., & Toinard, C. (2009) "Security and results of a large-scale high-interaction honeypot", *Journal of Computers*, Special Issue on Security and High Performance Computer Systems, Vol. 4, No. 5, pp. 395–404.

[3] Brunner, R. (2006) "A performance evaluation of the kad-protocol", Master's thesis, University of Mannheim and Institut Eurecom.

[4] Chaum, D. L. (1981) "Untraceable electronic mail, return addresses, and digital pseudonyms", *Communications of the ACM*, Vol. 24, No. 2, pp. 84–90.

[5] Grothoff, C., Grothoff, K., Horozov, T., & Lindgren, J. T. (2003) "An Encoding for Censorhip-Resistant Sharing", Technical report.

[6] Clarke, I., Miller, S. G., Hong, T. W., Sandberg, O., & Wiley, B. (2002) "Protecting free expression online with freenet", IEEE *Internet Computing*, Vol. 6 No. 1, pp. 40–49.

[7] Clarke, I., Sandberg, O., Wiley, B., & W.Hong, T. (2000) "Freenet: a distributed anonymous information storage and retrieval system", In International Workshop on Design Issues in Anonymity and Unobservability, pp. 311–320.

[8] Ernesto (2007) "Comcast throttles bittorrent traffic, seeding impossible", World Wide Web electronic publication.

[9] Stoica, I., Morris, R., Karger, D., Kaashoek, M. F., & Balakrishnan, H. (2001), "Chord: A scalable peer-to-peer lookup service for internet applications", Technical report, New York, NY, USA, pp. 149-160.

[10] Bennett, K. & Grothoff, K., (2002). GAP - pratical anonymous networking. Technical report, Departement of Computer Sciences, University of Purdue (USA).

[11] Bennett, K., Grothoff, C., Horozov, T., Patrascu I., & Stef, T. (2002) "GNUNET - A truly anonymous networking infrastructure", Technical report, Departement of Computer Sciences, Purdue.

[12] Kügler, D. (2003) "An analysis of Gnunet and the implications for anonymous, censorship-resistant networks", In Proceedings of the 3rd International Workshop on Privacy Enhancing Technologies, pp. 161–176, Springer-Verlag.

[13] Maymounkov, P., & Mazieres, D. (2002) "Kademlia: A peer-to-peer information system based on the XOR metric", In International Workshop on Peer-to-Peer Systems (IPTPS), LNCS, Vol. 1.

[14] Dingledine, R., Freedman, M. J., & Molnar, D. (2000) "The Free Haven Project : Distributed Anonymous Storage Service", Technical report, MIT and Harvard.

[15] Sandhu, R. S. (1994), "On five definitions of data integrity". In The IFIP WG11.3 Working Conference on Database Security VII, Amsterdam, The Netherlands: North-Holland Publishing Co, pp. 257–267.

[16] Sherman, C. (2000), "Napster: Copyright killer or distribution hero?", *Online*, Vol. 24, No. 6, pp. 6-28.

[17] Ratnasamy, S., Francis, P., Handley, M., Karp, R., & Schenker, S. (2001) "A scalable content-addressable network", In SIGCOMM'01: Proceedings of the 2001 conference on Applications, technologies, architectures, and protocols for computer communications, New York, NY, USA: ACM, pp. 161–172.

[18]     Wagner, A., Dubendorfer, T., Hammerle, L., & Plattner, B. (2006) "Flow-based identification of p2p heavy-hitters", In ICISP'06: Proceedings of the International Conference on Internet Surveillance and Protection, Washington, DC, USA: IEEE Computer Society, pp. 15.